\documentclass[]{elsart3p}
\usepackage{graphicx}
\journal{Physica D}

\begin{document}
\begin{frontmatter}

\title{Contraction-induced cluster formation in cardiac cell culture}
\author[Fukui]{Takahiro Harada\corauthref{ca}},
\corauth[ca]{Corresponding author}
\ead{harada@life.ne.his.fukui-u.ac.jp}
\author[Kyoto]{Akihiro Isomura},
\author[Kyoto]{Kenichi Yoshikawa}
\address[Fukui]{Department of Human and Artificial Intelligent Systems, University of Fukui, Fukui 910-8507, Japan}
\address[Kyoto]{Department of Physics, Graduate School of Science, Kyoto University, Kyoto 606-8502, Japan \& Spatio-temporal Order Project, ICORP, JST}

\begin{abstract}
Evolution of the spatial arrangement of cells in a primary culture of cardiac tissue derived from newborn rats was studied experimentally over extended period.
It was found that cells attract each other spontaneously to form a clustered structure over the timescale of several days.
These clusters exhibit spontaneous rhythmic contraction and have been confirmed to consist of cardiac muscle cells.
Addition of a contraction inhibitor (2,3-butanedione-2-monoxime) to the culture medium resulted in the inhibition of both the spontaneous contractions exhibited by the cells as well as the formation of clusters.
Furthermore, the formation of clusters is suppressed when high concentrations of collagen are used for coating the substratum to which the cells adhere.
From these experimental observations, it was deduced that the cells are mechanically stressed by the tension associated with repeated contractions and that this results in the cells becoming compact and attracting each other, finally resulting in the formation of clusters.
This process can be interpreted as modulation of a cellular network by the activity associated with contraction, which could be employed to control cellular networks by modifying the dynamics associated with the contractions in cardiac tissue culture.
\end{abstract}


\end{frontmatter}

\section{Introduction}
Primary cultures of cardiac muscle cells have been successfully utilized as experimental models to study the complex electrical and mechanical activity of the heart.
This is because, compared to experiments using an intact heart, a variety of methods can be employed with relative ease to visualize and modify the dynamical activity of cells.
Taken together, these advantages enabled us to explore the nature of the complex dynamics that arise in cell cultures \cite{Rohr:1997, Bub:1998, Bub:2002, Iravanian:2003, Bub:2003, Hwang:2005}.

Previous studies revealed that the spatio-temporal patterns in the cellular activity of rhythmic contraction in cardiac muscle cell cultures undergo qualitative changes depending on the age of the culture.
Soen {\it et al.} investigated the spontaneous contraction of a monolayer of cardiac cell culture for over one week, and found that various interesting temporal patterns, including subharmonic and alternant rhythms, arise spontaneously \cite{Soen:1999}. Spontaneous emergence and annihilation of spiral waves were also observed in several studies \cite{Bub:1998, Bub:2002, Hwang:2004}. Kawahara and coworkers reported that formation of cellular network decreases the fluctuations in the timings of contraction \cite{Kawahara:2002, Kawahara:2004}.
One possible reason for this phenomenon is an alternation in the degree of gap junction formation which brings a passive conductance between cells \cite{Oyamada:1994, Meiry:2001}.
Interestingly, despite having been implicated in affecting the characteristics of intercellular communication, relatively little attention has been directed at the evolution of the spatial arrangement between cells.
This may be due to the current belief that cardiac muscle cells, because they differ from self-propelling cells, including fibroblasts, do not possess the molecular apparatus necessary for facilitating active migration.
However, even if a single cardiac muscle cell is unable to migrate by itself, it may be possible for the geometrical arrangement of the cells in a population to change via interactions among the cells.

To more accurately assess such a possibility, we investigated the evolution of the spatial arrangement of cells in a cardiac cell culture over an extended timescale in the present study.
We found that cardiac muscle cells spontaneously aggregated and formed clusters in which cells constituting the same cluster periodically contracted in synchrony.
The application of an agent that inhibits the contraction of cells results in the suppression of cluster formation, suggesting that the periodic contraction of individual cardiac muscle cells affects cluster formation.
Furthermore, the degree of cluster formation depends on the concentration of extra cellular matrix to which the cells adhere.
These experimental findings indicate that the tension generated by periodic contraction of cardiac muscle cells contributes markedly to the driving force underlying cluster formation.

The results of the present study suggest that the rhythmic contraction of cardiac muscle cells affects the spatial arrangement and structure of the cellular network.
This phenomenon is likely to be concerned with age-dependent changes in the spatio-temporal dynamics of the electric activity of cells observed in earlier studies.
This new hypothesis, contraction-induced modulation of an intercellular network, might provide a new class of dynamical phenomena in a cardiac tissue culture.

\section{Materials and Methods}
\subsection{Preparation of cell culture}
Primary cultures of ventricular cells were prepared from one-day-old newborn rats according to methods described elsewhere \cite{Harada:2006, Matoba:1999}.
Briefly, ventricles isolated from the hearts of one-day-old newborn rats were minced and were treated with collagenase.
Isolated cells were collected by centrifugation and were supplied with plating medium (Dulbecco-modified Eagle medium containing 10\% fetal bovine serum, 1\% penicillin/streptomycin and 0.1 mM 5-bromo-2'-deoxyuridine (BrdU)).
In order to reduce the population of non-muscle cells, the method of selective plating was employed \cite{Harada:2006, Matoba:1999}.
Then, the cells were counted and diluted to desired densities (from $2 \times 10^5$ cells/ml to $6\times 10^5$ cells/ml).
Because accurate estimation of the density of cells at this stage is technically difficult, not the absolute values but the relative values of initial cell density given in the following sections have a meaning.
The cells were plated on the collagen-coated petri dishes (see below), which was followed by incubation for 24 h.
Subsequently, the medium was exchanged with contraction medium (modified Eagle medium with 10\% calf serum, 1\% penicillin/streptomycin, and 0.1 mM BrdU), and the cells were incubated under the same conditions as described above.

\subsection{Preparation of collagen-coated culture dish}
Polystyrene petri dishes (BD Biosciences, CA) coated with collagen were used for experiments.
In experiments described in Sections~\ref{ss.clustering} to \ref{ss.contraction}, dishes were coated with collagen by incubating them with collagen (Type I, derived from rat tail, WAKO, Osaka) solubilized in acetic acid (0.1 \%, pH 3) for several minutes at room temperature.
In experiments described in Section~\ref{ss.substratum}, the amount of collagen used to coat dishes was controlled as follows:
First, varying concentrations of collagen solutions (1.2 mg/$\ell$ to 120 mg/$\ell$) diluted in phosphate-buffered saline (PBS) were applied on petri dishes, which were then incubated at 4$^\circ$C for 7 h. Dishes were washed with PBS several times before use.

\subsection{Time-lapse observations}
In order to observe the process of cluster formation, time-lapse observation method was adopted.
After exchanging medium on the first day, a culture dish was positioned on an on-stage incubation chamber (IN-0NI-F2, Tokai Hit, Shizuoka) and examined under an inverted microscope (IX-70, Olympus, Tokyo). The culture was maintained under humidified conditions at 37$^\circ$C and 5\% CO$_2$.
The cell culture was observed using a 10$\times$ objective lens under bright-field illumination and images were captured every 3 min for up to a week, using a digital charge-coupled-device (CCD) camera (C4742-95-12SC, Hamamatsu Photonics, Shizuoka). The culture medium was not exchanged during the time-lapse observations.

\subsection{Determination of beat-rate distributions}
The spatial distribution of the beat rate of cardiac muscle cells was studied as follows. Culture dishes were incubated at the same condition as described above. Spontaneous contraction of cardiac muscle cells were observed every 24 h using inverted phase-contrast microscope (4$\times$ magnification) and recorded at 30 frames per seconds for 3 minutes under the room temperature (25 $^\circ$C).
The sequential images (960 $\mu$m$\times$ 960 $\mu$m) were subdivided into small segments of 40$\mu$m$\times$ 40 $\mu$m (the size roughly corresponding one or two cells), for each of which average gray value was calculated to yield a set of time series. In these time series, clear peaks associated with spontaneous contractions of cells in each segment occur \cite{Soen:1999}. Thus, the number of peaks in one minute was counted in order to calculate the beat rate of cells in each segment.

\subsection{Determination of cellular orientation}
The orientational order of cells was investigated, following the procedure of Ref.~\cite{Kemkemer:2000}, using bright-field time-lapse images.
First, an image was subdivided into segments of 15 $\mu$m$\times$ 15 $\mu$m, and a Fourier power spectrum was calculated for each segment.
The cross-correlation matrix of each Fourier spectrum was calculated, and the principal axes were determined.
It has been known that the principal axis with the smaller eigenvalue represents the mean direction of cellular arrangement \cite{Kemkemer:2000}.
The difference between the two eigenvalues can be interpreted as the degree of orientational order, because it increases as the anisotropy of the original image increases.

\subsection{Histochemical analysis}
Cells were stained with a monoclonal antibody, 5C5, directed against $\alpha$-sarcomeric actin (A2172, SIGMA, Tokyo).
Furthermore, Vectashield (Vector Laboratories, U.~K.) was used to overlay the cultures to prevent the sample from drying out.
Vectashield also contains 4',6-diamino-2-phenylindole (DAPI) which stains the nuclei of the cells.
The stained cells were observed with an epifluorescent inverted microscope (IX-70, Olympus, Tokyo), and images were captured using a digital CCD camera (DV887ECS-UVB, Andor Technology, Northern Ireland).

\subsection{Analysis of cell distribution}
The spatial distribution of cell nuclei was analyzed to assess the spatial distribution of cells.

% Large scale
First, the distribution of the cellular clusters were determined as follows. At the seventh day from the onset of incubation, the cells in culture were fixed using cold methanol (-20 $^\circ$C).
The phase-contrast images of the culture were obtained for 16 $\times$ 16 fields of view, each of which has the size of 880 $\mu$m$\times$ 704 $\mu$m and is 1 mm apart from its neighbors, using a motorized mechanical stage (BIOS-105T, Sigma Koki, Tokyo).
The number of clusters was counted manually from these images.

% Small scale
Second, the distribution of the cells was determined at a higher resolution as follows.
The fluorescent images of cell nuclei stained with DAPI were obtained for at least 79 fields of view (820 $\mu$m $\times$ 820 $\mu$m each) for every culture dish. At least 9700 nuclei are included in the analysis for each sample.
No significant difference among the samples in the same condition was observed in the essential trends in the results of following analysis.
The images were processed and analyzed with image processing software ImageJ (a public domain software originally developed at the US National Institute of Health).
The images were converted to binary images by adopting a manually determined threshold of gray value.
Because there is a tendency that the images of neighboring cell nuclei become indistinguishable in the region with high cellular density, i.e., in clusters, the watershed algorithm was applied in order to separate them \cite{Russ:1995}, although the separation was not complete.
This leads to underestimation of the cellular density in the vicinity of clusters.

On the basis of the coordinates of the identified nuclei, a radial distribution function for cell nuclei was calculated as
\begin{equation}
g(r) \equiv \frac{n(r)}{2\pi r},
\end{equation}
where $n(r) \mathrm{d}r$ denotes the average number of nuclei falling within a distance of $(r, r + \mathrm{d}r)$ from a reference nucleus.
In order to avoid the boundary effect, in each field of view, the reference cells were selected from a region distant from a certain distance $R$ from the boundary of the image ($R = 200 ~\mu$m).
$g(r)$ has the unit of number density, and is normalized so that its value at a sufficiently large $r$ gives the average number density of nuclei. 

Third, the distribution of the size of clusters was determined as follows.
For each field of view, cellular nuclei whose center is within 50 $\mu$m from the center of another nuclei were considered to be included in the same cluster.
Then, the number of cells constituting the same cluster was calculated for all the fields of view.

\subsection{Analysis of contractile motion of cells} \label{ss.move}
In order to characterize cellular contraction, motion analysis of fluorescent microbeads attached to the surface of the cells was conducted.
The culture medium was exchanged on the sixth day of culture with contraction medium containing $2 \times 10^{-4}$ \% (v/v) of fluorescent polystyrene beads (Molecular Probes, 1 $\mu$m diameter).
After one hour of incubation, these beads were sedimented in the culture dish and became uniformly attached to the cell surface.
The density of the beads was calculated to be $2.4 \times 10^{3}$/mm$^2$ (approximately six beads per cell).
The beads were found to move in phase with the contractions exhibited by each cardiac muscle cell.
The motion of the fluorescent beads was captured using an intensified CCD camera (C2400-27, Hamamatsu Photonics, Shizuoka).
Using ImageJ software, the proportion of beads with centers of mass that had trajectories exceeding 3.0 $\mu$m were calculated.

\section{Results}
\subsection{Formation of clusters} \label{ss.clustering}

\begin{figure}[tbp]
\begin{center}
\includegraphics{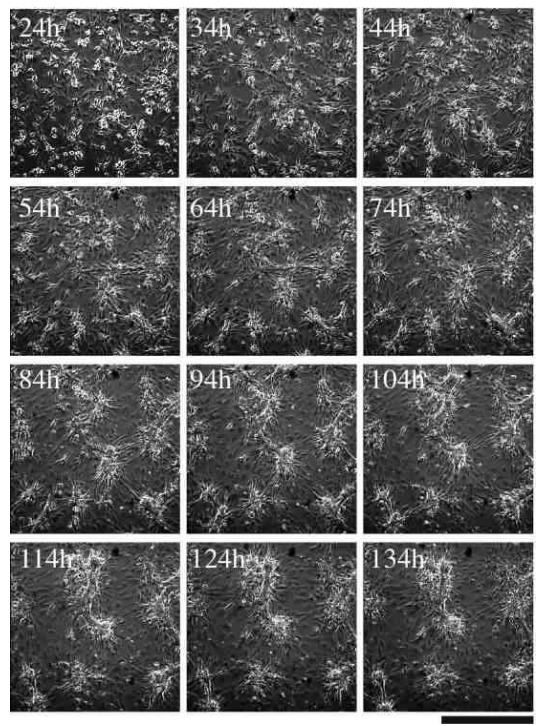}
\caption{Changes in the arrangement of cardiac cells over time.
The image sequence shows the changes in the cell culture observed from the onset of incubation to 134 h. The scale bar represents 500 $\mu$m.
}
\label{f.timecourse}
\end{center}
\end{figure}

First, the cells were plated at an initial density of $4.4 \times 10^2$ cells/mm$^2$.
Changes in the arrangement of cells over time was recorded using time-lapse observation methods; as shown in Fig.~\ref{f.timecourse}, the cells exhibited a variety of dynamic organization structures.

\begin{figure}[tbp]
\begin{center}
\includegraphics{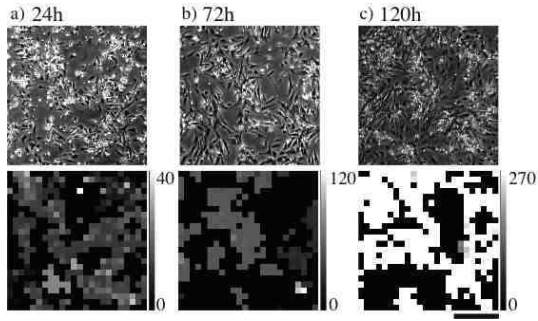}
\caption{Changes in the beat rate of spontaneous contraction through development of clusters. The upper images display the phase-contrast microscopic images at a) 24 h, b) 72 h, c) 120 h from the onset of incubation. These images were captured for the different samples.  The scale bar represents 300 $\mu$m. The lower images represent the beat rates of spontaneous contraction for the same fields of view as the upper images. Beat rates (beats per minutes) were displayed using the gray level indicated on the right of each figure.
}
\label{f.dynamics}
\end{center}
\end{figure}

Initially, (24 h to 44 h from the onset of incubation), the distribution of cells was approximately uniform, but the number of the cells that adhered to the substrate increased with time.
At this stage, a part of the cells started exhibiting spontaneous contractions.
However, the frequencies and phases of periodic contraction were observed to vary among cells, and synchronization of periodic contraction is restricted to small aggregations of cells (see Fig.~\ref{f.dynamics}a).
After approximately 54 h of incubation, small clusters of cells became distinct.
As aggregation of the cells proceeded, they formed a network of interconnected small clusters (74 h to 94 h).
At this stage, the cells that formed aggregations were found to contract periodically, and the cells belonging to the same cluster tended to exhibit a synchronized contractile motion; however, global synchronization of the timings of periodic contraction was not observed (Fig.~\ref{f.dynamics}b).
Aggregation of cells continued, with the interconnected network of small clusters forming several distinct clusters of cells (after 124 h), and acquiring a round shape (134 h).
At this stage, spontaneous contraction was only observed in the large, developed clusters, whereas the cells outside clusters did not exhibit any rhythmic contractions.
The periodic contractions of the large clusters were globally synchronized throughout the culture dish (Fig.~\ref{f.dynamics}c).

\begin{figure}[tbp]
\begin{center}
\includegraphics{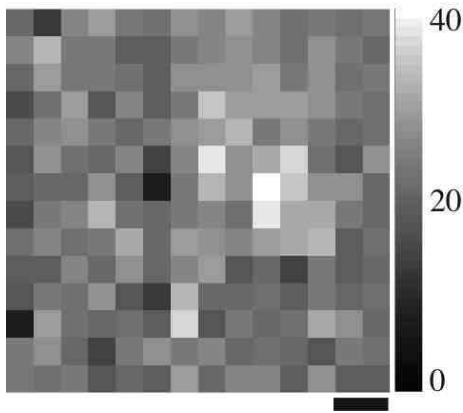}
\caption{
Number density of clusters on the seventh day from the onset of incubation. The scale bar represents 2 mm. The number density (number of clusters per mm$^2$) is represented as a gray value indicated at the right of the image.
}
\label{f.global}
\end{center}
\end{figure}

As illustrated in Fig.~\ref{f.global}, the distribution of clusters in the culture dish was almost uniform, although there exist an inhomogeneity caused by the initial inhomogeneity of cells. At this initial density of cells ($4.4 \times 10^2$ cells/mm$^2$), the number density of clusters was $22 \pm 5$ /mm$^2$ (Mean $\pm$ S.~D.).

\begin{figure}[tbp]
\begin{center}
\includegraphics{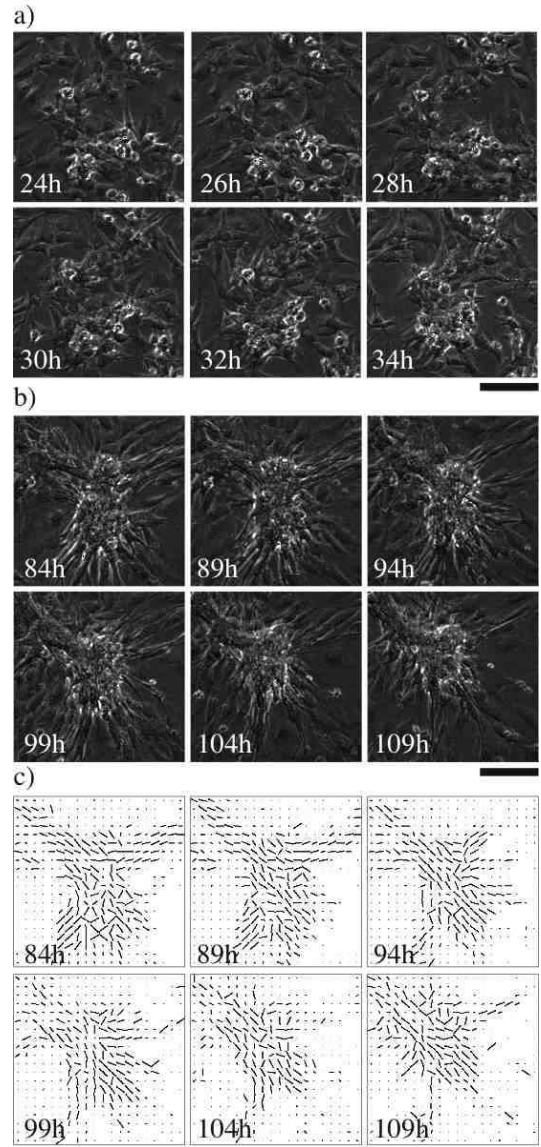}
\caption{Process of cluster formation at a higher resolution both on space and time for the same experiment as in Fig.~\ref{f.timecourse}. The scale bar represents 100 $\mu$m.
a) The early stage of cluster formation. The time from the onset of incubation is displayed on the each image (see also the supplementary movie).
b) The later stage of cluster formation (see also the supplementary movie).  c) Orientation of the cells is represented by solid lines for the same field of view as in b). The angle of each line represents the direction of the principal axis with the smaller eigenvalue of the cross-correlation matrix of the local Fourier transform. The length of each line is proportional to the difference between two eigenvalues, which therefore represents the degree of orientational order.
}
\label{f.director}
\end{center}
\end{figure}

Fig.~\ref{f.director}a displays the process of cluster evolution in the early stage at a higher spatio-temporal resolution than that shown in Fig.~\ref{f.timecourse}.
As seen in this figure, the parts with higher initial density serve as nuclei of cluster formation.
Fig.~\ref{f.director}b also displays the images of higher spatio-temporal resolution for the later stage of cluster formation.
For the same field of view, the field of orientational order of cells, which is obtained from the local Fourier transform of the images in Fig.~\ref{f.director}b, is represented in Fig.~\ref{f.director}c.
As observed in this figure, the clusters shrink into the direction in which they are oriented.
It has been known that the muscle fibrils, which are the cellular apparatus responsible for inducing contraction, tend to align along the long axis of the cell and contraction occurs in this direction.
The observation given in Fig.~\ref{f.director}b and c thus imply that the contraction of cells is related to the process of cluster formation and this possibility is discussed below.

\subsection{Identification of cell type} \label{ss.immuno}
Given that a primary culture was employed in the present study, the culture was contaminated by non-muscle cells, such as fibroblasts.
The cells outside the clusters appeared to be non-muscle cells, because while they exhibited active locomotion and proliferation, but no contraction.
We therefore undertook a histochemical analysis to determine which types of cells were involved in the formation of clusters.

\begin{figure}[tbp]
\begin{center}
\includegraphics{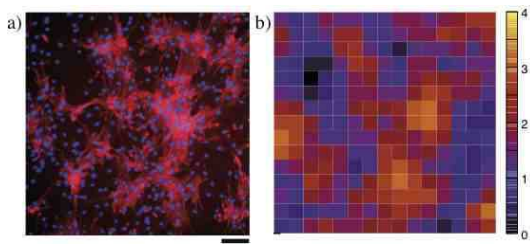}
\caption{a) Fluorescent image of cardiac cell culture on the sixth day after the onset of incubation represented in pseudocolor. $\alpha$-sarcomeric actin is colored red, while the cell nuclei are shown in blue. The scale bar represents 100 $\mu$m. b) Local density of nuclei calculated for the same field of view as shown in a). Density, displayed as a color map, is represented as the number of nuclei in each small block ($6.6 \times 10^2$ $\mu$m$^2$).}
\label{f.immuno}
\end{center}
\end{figure}

$\alpha$-sarcomeric actin (actin expressed in muscle cells) and cellular nuclei were stained using $\alpha$-sarcomeric actin antibody and DAPI (a dye that stains DNA), respectively.
Fig.~\ref{f.immuno}a shows a fluorescent image of cells at the sixth day from the onset of incubation.
Since $\alpha$-sarcomeric actin is specifically expressed in cardiac muscle cells, cell nuclei colocalized with $\alpha$-sarcomeric actin are considered to be cardiac muscle cells, while those cell nuclei that are not associated with $\alpha$-sarcomeric actin  are not muscle cells. Therefore, using this method and the associated images, we are able to discriminate between cardiac muscle cells and non-muscle cells.
The fraction of cardiac muscle cells, in which the nucleus colocalizes with $\alpha$-sarcomeric actin, was 0.76 $\pm$ 0.1.

It was also found that the cell clusters mainly consisted of cardiac muscle cells and that non-muscle cells were located outside the clusters.
Actually, the distribution of nucleus density (Fig.~\ref{f.immuno}b) was similar to that of $\alpha$-sarcomeric actin.
In regions where the local density of nuclei was large, which corresponded to clusters, the expression of $\alpha$-sarcomeric actin was evident.
From this observation, it is obvious that the clusters consist of cardiac muscle cells.
While there were nuclei outside clusters, these nuclei were not colocalized with $\alpha$-sarcomeric actin, which indicated that these cells were non-muscle cells.
The density of these cells was low and a region without cells was observed outside the clusters.

\subsection{Radial distribution of cells and the size of clusters}
\begin{figure}[tbp]
\begin{center}
\includegraphics{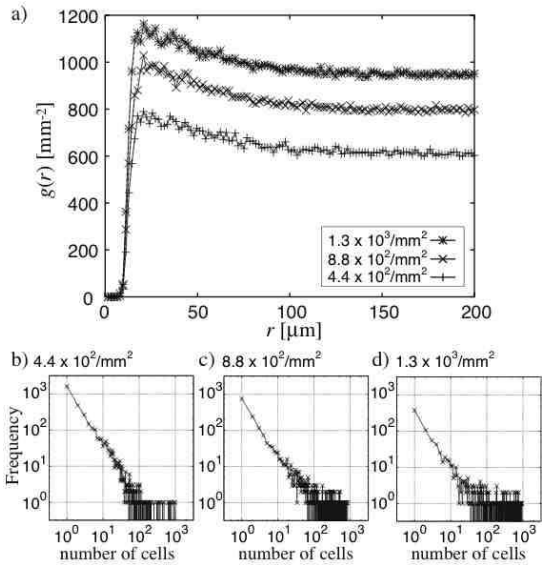}
\caption{Distribution of cell nuclei. a) Radial distribution function of cell nuclei for several initial cell density. The initial cell densities are indicated in the inset. Dotted curves are the fits to Eq.~(\ref{e.fit}).
b) - d) Histograms of number of cells in a single cluster. Initial cell densities are b) $4.4 \times 10^2$ cells/mm$^2$, c) $8.8 \times 10^2$ cells/mm$^2$, d) $1.3 \times 10^3$ cells/mm$^2$, respectively. The average numbers of cells in a cluster are a) 7.8 $\pm$ 23, b) 23 $\pm$ 124, and c) 40 $\pm$ 431, respectively.}
\label{f.dist}
\end{center}
\end{figure}

In order to quantify the distribution of cells, the cells were fixed on the seventh day, and the nuclei of cells were stained with DAPI. The images of the cell nuclei were obtained using fluorescent microscopy.
Figure~\ref{f.dist}a shows the radial distribution function of cell nuclei calculated from these images for several initial cell densities.
For all densities studied here, a peak is observed around 20 $\mu$m in the radial distribution function.
Considering the size of a nuclei and a cell (around 10 $\mu$m and a few tens $\mu$m, respectively), this peak is interpreted as the distance between nuclei of closely packed cells, i.e., the cells in the same cluster.
From Fig.~\ref{f.dist}a, we considered that the nucleus within 50 $\mu$m from each other constitute a single cluster. Using this definition, the number of nucleus (cells) inside a single cluster was determined for every cluster. Figure \ref{f.dist}b, c, and d displays the number distribution of nuclei in a cluster for several initial cell densities.
As seen in these figures, the number in a single cluster varies from one to a thousand. As the initial density of cells increases, the fraction of large (e.g., $\ge$ 10 nuclei) clusters increases.

\subsection{Effect of cell contraction} \label{ss.contraction}
To examine the effect of spontaneous cardiac muscle-cell contraction on cluster formation, we investigated whether inhibition of contraction affected cluster formation.
For this purpose, we utilized 2,3-butanedione 2-monoxime (BDM), which is known to affect the actomyosin system in muscle fibrils and inhibit contraction \cite{Li:1985, Fujita:1998}.

\begin{figure}[tbp]
\begin{center}
\includegraphics{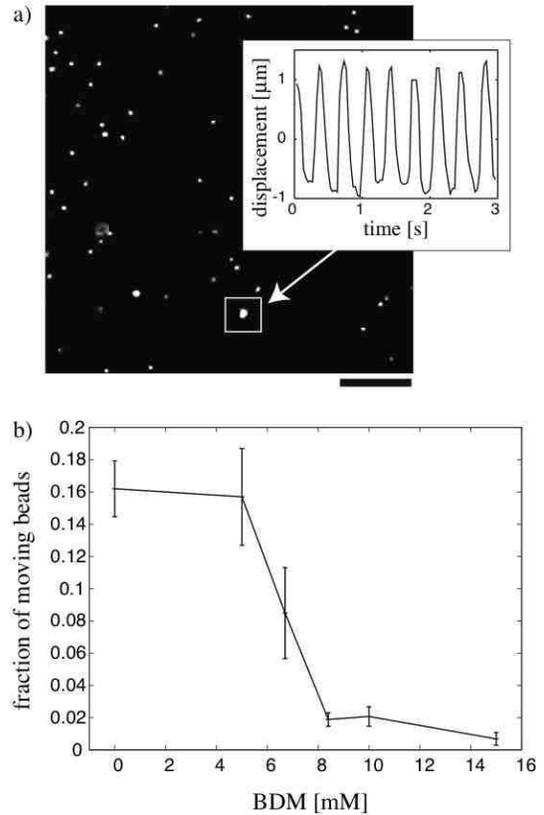}
\caption{Effect of BDM on the spontaneous contraction of cardiac muscle cells. a) Moving beads assay. Fluorescent image of microbeads ($\phi$ 1 $\mu$m) attached to the surface of cells supplied with culture medium containing no BDM. Inset exemplifies the motion of the bead that is marked in the image. The scale bar represents 50 $\mu$m. b) The fraction of moving fluorescent beads attached to cell surface is plotted against the concentration of BDM. The error bars represent $\pm$S.E.M.}
\label{f.move}
\end{center}
\end{figure}

Application of this agent to culture medium resulted in the inhibition of spontaneous contraction of cells.
The activity of spontaneous cellular contraction was estimated using the method described in Section~\ref{ss.move}.
In this method, fluorescent microbeads ($\phi$ 1 $\mu$m) were dispersed on the surface of the cells and the fraction of periodically moving beads was determined using video microscopy.
Figure~\ref{f.move} depicts the fraction of periodically moving beads with varying concentration of BDM.
As illustrated in this figure, the fraction of periodically moving beads decreases as the concentration of BDM increases. In the all preparations, the presence of 10 mM of BDM almost completely inhibited the spontaneous contraction of cardiac muscle cells.
It has been confirmed that the cells were alive at these concentrations of BDM, because the spontaneous contraction resumed after removing the BDM and exchanging the culture medium with fresh medium without BDM (data not shown).

\begin{figure}[tbp]
\begin{center}
\includegraphics{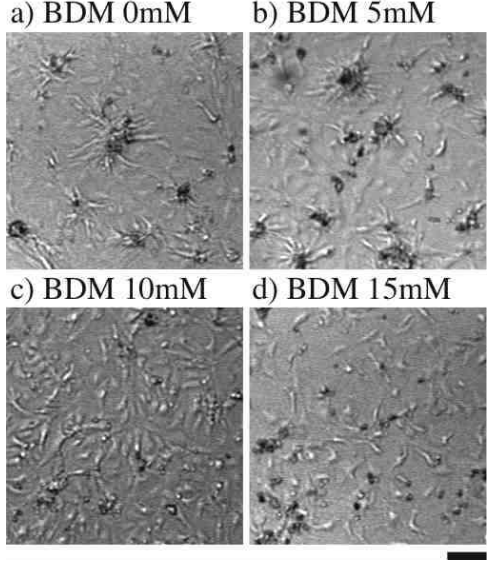}
\caption{Microscopic images of cells on the fifth day from the onset of incubation in the presence of various BDM concentrations. The concentration of BDM is indicated on the top of each figure. The scale bar represents 100 $\mu$m.}
\label{f.bdm}
\end{center}
\end{figure}

\begin{figure}[tbhp]
\begin{center}
\includegraphics{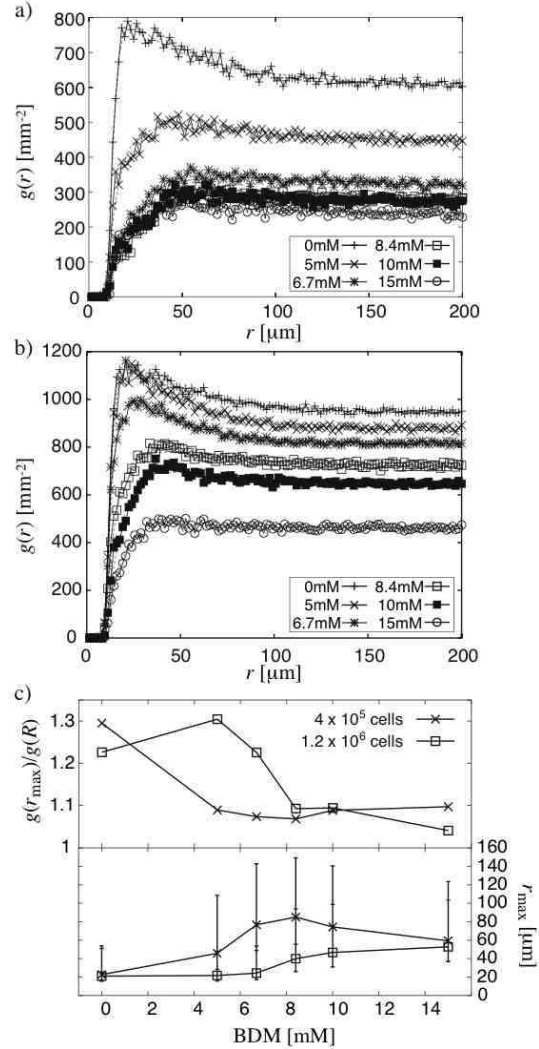}
\caption{Radial distribution functions of cell nuclei on the seventh day. Different curves represent the radial distribution function at different concentrations of BDM. a) Initial density of cells was $4.4 \times 10^2$ cells/mm$^2$. b) Initial density of cells was $1.3 \times 10^3$ cell/mm$^2$. c) Intensity (top) and the location (bottom) of the peak radial distribution function.
Experimental data displayed in a) and b) were fitted to Eq.~(\ref{e.fit}), from which the intensity and the location of the peak was estimated.
In a) and b), the functions, $\bar g(r)$, obtained by fitting are also represented using dotted curves.
The intensity is represented as the ratio $\bar g(r_{\rm max})/\bar g(R)$, with letting $r_{\rm max}$ be the location of the peak and $R = 200$ $\mu$m.
In the bottom plot, the error bars indicate the half width of $\bar g(r) - \bar g(R)$. Crosses and squares curves represent data derived from the plots in a) (lower density) and b) (higher density), respectively.}
\label{f.rdf}
\end{center}
\end{figure}

The cells were supplied with culture medium containing varying concentrations of BDM and incubated for several days.
Figure~\ref{f.bdm} shows the microscopic images of cells on the fifth day of culture. This figure illustrates that the formation of cellular clusters is suppressed in the presence of a high concentrations of BDM ($\ge$ 10 mM).
In order to quantify the distribution of cells, the cells were fixed on the seventh day, and the nuclei of cells were stained with DAPI. The images of the cell nuclei were obtained using fluorescent microscopy.
Figure~\ref{f.rdf}a shows the radial distribution function of cell nuclei calculated from these images.
For cells cultivated in the absence of BDM, the radial distribution function exhibits a peak at approximately 20 $\mu$m, indicating the formation of clusters.
As the concentration of BDM increases, the intensity of the peak decreases and the location of the peak shifts towards a larger distance.
At 15 mM of BDM, the radial distribution function exhibits almost no peaks, indicating the absence of clusters and a random distribution of cells.

In order to confirm the abovementioned trends quantitatively, we determined the intensity and the location of the peaks.
Since the raw data is rather noisy, the experimental data were fitted to the following function
\begin{equation}
\bar g(r) = A\exp\left[-B \left(\frac{r}{d}\right)^{-n} + C \left(\frac{r}{d}\right)^{-m}\right],
\label{e.fit}
\end{equation}
from which we determined the intensity and the location of the peaks.
As observed in Fig.~\ref{f.rdf}, this function fits the experimental data well, and we can estimate several parameters from the fitting function.
In Fig.~\ref{f.rdf}c, the intensity and the location of the peak are plotted as functions of BDM concentration.
The intensity of the peak is represented as the ratio of the magnitude of the radial distribution function at the peak to the magnitude of the radial distribution function at a large distance $R$. In this case, we selected $R = 200$ $\mu$m.
Based on these plots, it became clear that the intensity of the peak decreases and the location of the peak shifts towards a larger distance as the concentration of BDM is increased.

One may notice that the average density of nuclei decreases as the concentration of BDM increases in Fig.~\ref{f.rdf}a.
This could be because the number of dead cells increases as the concentration of BDM in culture medium increases.
However, the disappearance of the peak in the radial distribution function with an increase in the BDM concentration is not simply attributable to a decrease in the average density of cells.
In fact, the morphology of cells is clearly different for different concentration of BDM (see Fig.~\ref{f.bdm}).
Moreover, when the initial density of cells was increased three fold, a similar trend regarding was observed between the concentration of BDM and the shape of the radial distribution function (see Fig.~\ref{f.rdf}b and Fig.~\ref{f.rdf}c).
The change of the shape of the radial distribution function induced by application of BDM is qualitatively different from a mere shift of the ordinate caused by the change in the cell density (see Fig.~\ref{f.dist}).

Interestingly, a marked correspondence was observed between spontaneous contraction activity and the structure of the radial distribution function regarding the dependence of the contractions on the concentration of BDM.
At a BDM concentration of less than 6.7 mM, both the contraction of cells and the formation of clusters were evident, while at a BDM concentration greater than 8.4 mM, both the contraction of cells and the formation of clusters were suppressed (see Figs.~\ref{f.move} and \ref{f.rdf}c).
Based on this observation, it became apparent that the spontaneous contraction of cardiac muscle cells was closely related to the formation of clusters.

\subsection{Effect of substratum} \label{ss.substratum}
Next, we examined the effect of the adhesive force of the muscle cells to the substratum on the formation of clusters.
In the abovementioned experiments, substrata coated with a thin layer of collagen were used.
Given that the cells were attached to the substratum by a collagen coating, the effect of the strength of cellular adhesion to the substratum was tested by altering the amount of collagen in the coating.

\begin{figure}[tbp]
\begin{center}
\includegraphics{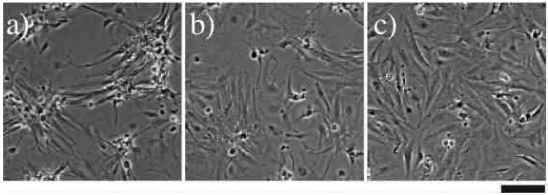}
\caption{Microscopic images of cells on the substratum coated with varying concentrations of collagen solution.
The images of cells fixed on the seventh day from the onset of incubation are displayed.
The scale bar represents 100 $\mu$m.
The concentrations of collagen utilized to coat substrata are 1.2 mg/$\ell$ (a), 3.6 mg/$\ell$ (b), 120 mg/$\ell$ (c).}
\label{f.coll}
\end{center}
\end{figure}

\begin{figure}[bp]
\begin{center}
\includegraphics{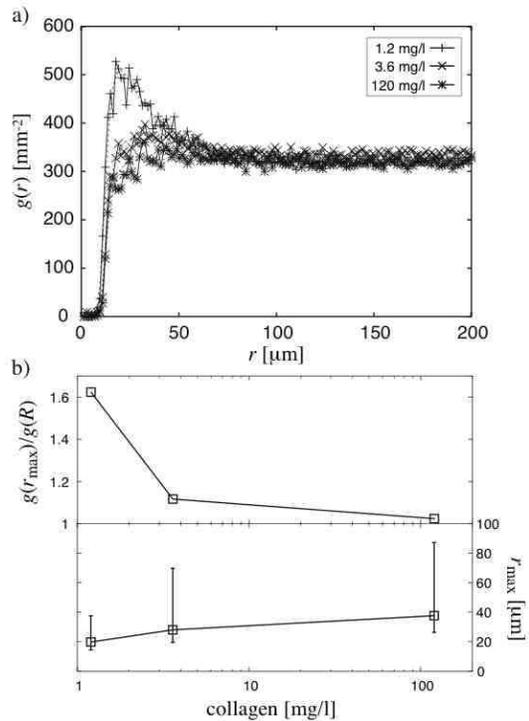}
\caption{a) Radial distribution function of cell nuclei on the seventh day. Different curves represent the radial distribution function at the different collagen concentrations used to coat the substrata. b) Parameter values obtained from fitting the curves in a) by Eq.~(\ref{e.fit}). The functions, $g(r)$, obtained by fitting are represented in a) using dotted curves. (Top) The intensity of the peak defined as $\bar g(r_{\rm max})/\bar g(R)$, where $R = 200$ $\mu$m. (Bottom) The location of the peak plotted as a function of concentration of collagen. Error bars indicate the half width of $\bar g(r) - \bar g(R)$.}
\label{f.rdf_c}
\end{center}
\end{figure}

We then prepared substrata coated with collagen at varying concentrations.
The cells were plated on these substrata and incubated for several days under the same conditions as those described above. It was confirmed by visual inspection that the cells exhibited apparent contractions independent of the concentration of collagen.
Figure~\ref{f.coll} shows microscopic images of cells plated on substrata coated with collagen at varying concentrations. Interestingly, clusters formed when the concentration of collagen was low. Conversely, no clusters formed when the concentration of collagen was high.
In order to quantify this observation, the radial distribution function was calculated.
On the seventh day from the onset of incubation, the cells were fixed and stained with DAPI.
The distribution of cell nuclei was then investigated using fluorescent microscopy, and the radial distribution function was calculated.
Figure~\ref{f.rdf_c}a displays the radial distribution functions of cultures grown on substrata with varying amounts of collagen.
As seen in this figure, when the concentration of collagen was low (1.2 mg/$\ell$), the radial distribution function exhibits a clear peak around 20 $\mu$m, which is similar to the former experiments in the absence of BDM.
Furthermore, the peak in the radial distribution function became less intense and broad as the concentration of collagen increased. In substrata with a high concentration of collagen (120 mg/$\ell$), almost no peak was observed in the radial distribution function. These observations were also confirmed by fitting Eq.~(\ref{e.fit}) to the radial distribution functions, which yielded estimates of the intensity and the location of the peaks (see Fig.~\ref{f.rdf_c}b).
Consequently, while the formation of clusters was suppressed when the substratum was coated with high concentrations of collagen, cells were able to form clusters when collagen concentrations in the substratum were low.

\section{Discussion}
In the present study, spontaneous formation of cell clusters was observed in primary cultures of ventricular cells from newborn rats.
While the formation of clusters has been observed in several earlier studies using similar preparations \cite{Lokuta:1994, Simpson:1994}, the mechanisms underlying cluster formation have not been clarified. 
In the present study, we have obtained several new insights into the mechanism of cluster formation.

First, as revealed by histochemical analysis, clusters were comprised primarily of cardiac muscle cells, with non-muscle cell types, such as fibroblasts, being located outside the clusters.

Second, the cells in a cluster are almost fully packed. The number of cells incorporated in the same cluster increases as the density of cells increases.

Third, the direction in which the cells move is correlated to the direction of the long axis of cells and the direction of cell contraction (Fig.~\ref{f.director}).
Furthermore, the observation that the inhibition of spontaneous contraction of cardiac muscle cells resulted in suppression of cluster formation (Figs.~\ref{f.bdm} and \ref{f.rdf}) supports the idea that the spontaneous contraction of cardiac muscle cells is related to cluster formation. It is thus plausible that the periodic contraction generated tension between the cells, and that this tension makes the cells move.

Fourth, the concentration of the collagen coated on the substratum affected cluster formation.
When the substratum was coated with a solution containing a high concentration of collagen, formation of cell clusters was suppressed (Figs.~\ref{f.coll} and \ref{f.rdf_c}).
The adhesive force of cells to the substratum is expected to increase when the concentration of collagen used to coat the substrate increases, because the number of contacts formed between the cytoskeleton and collagen would increase.
Consequently, cluster formation is suppressed when the adhesive force between cells and the substratum is large.

\begin{figure}[tb]
\begin{center}
\includegraphics{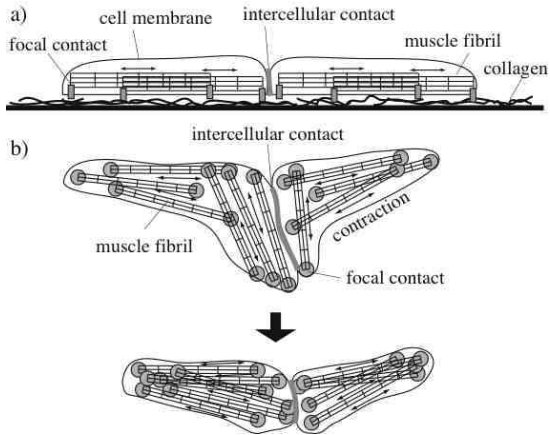}
\caption{Hypothetical mechanism of cluster formation. a) Schematic cross section of a cardiac muscle cell. Cells contain muscle fibrils, which become connected to collagen via focal contacts at their ends. The substratum is coated with collagen. Contractile motion is indicated by arrows and adjacent cells are connected through intercellular contacts, including adherence junctions and desmosomes. b) Dorsal view of two cardiac muscle cells in contact with each other via an intercellular contact. Individual cells contain muscle fibrils and exhibit periodic contractions, resulting in  the focal contacts at the ends of the muscle fibrils experiencing periodic force. These actions lead to the rearrangement of focal contacts and cytoskeletons in individual cells, which brings the two cells closer together.}
\label{f.scheme}
\end{center}
\end{figure}

Based on the above observations, the mechanism of cluster formation by cardiac muscle cells is proposed, as schematically shown in Fig.~\ref{f.scheme}.
In the first stage, the distribution of cells is random. At the initial cell density ($4.4 \times 10^2$ cells/mm$^2$), the area of substratum covered by cells is approximately half. Consequently, individual cells are likely to being in contact with several neighboring cells.
Once two cardiac muscle cells establish contact, gap junctions are formed \cite{Oyamada:1994}, and their contractions become synchronized (see Fig.~\ref{f.dynamics}).
These cells are also mechanically coupled via intercellular contact, including adherence junctions and desmosomes \cite{Atherton:1986}.
Contraction of a cell is attributed to muscle fibrils, which are complexes of cytoskeletons. The contraction of the muscle fibrils exerts tension on cytoskeletons, including actin filaments, which are anchored to the substratum via molecular apparatuses termed focal contact (see Fig.~\ref{f.scheme}a).
Owing to this tension, focal contacts at the periphery of cells are pulled into the center of each cell (this process may involve rapid destructions and reconstructions of focal contacts).
Through this process, cytoskeletons are rearranged and become compact.
Intercellular contacts are also expected to play an important role in maintaining and stabilizing the mechanical contacts between cells, against the contractile tension of cytoskeletons \cite{Toyofuku:2000}.
If these intercellular connections are sufficiently tight, the center of each cell become closer to each other because each cell becomes compact (see Fig.~\ref{f.scheme}b).
Since a number of cells are mechanically connected to each other and each cell becomes compact, the population of cells becomes denser and forms a cluster.

If the force generated by contraction is too weak to move focal contacts, contraction of cells will not result in the cells moving closer.
Similarly, if the concentration of collagen is large, the number and the strength of focal contacts between the cytoskeleton and collagen would increase, making rearrangement of the cytoskeleton less likely and leading to the suppression of cluster formation.
Furthermore, as the density of cells increases, the number of cells that are initially in contact increases, resulting in the larger size of clusters.
This scenario is consistent with the experimental results on the effects of BDM and also of the collagen concentration.

With regard to the cluster formation in cardiac cell culture, it has been reported previously that the presence of fibroblasts is important for inducing the aggregation of cardiac muscle cells in some situations \cite{Toyofuku:2000}.
In the present study, given the difficulties associated with complete removal of proliferating non-muscle cells from cultures, it is not clear whether the presence of non-muscle cells, whose fraction was not large in the present study (around 20 \% even in the late stage of culture), is essential for the observed phenomena.
However, as discussed above, several observations in the present study suggest that formation of clusters is also mediated by the interactions among cardiac muscle cells.

Finally, let us discuss the relation of the present results to the previous studies on the wave dynamics on the cardiac tissue culture.
The formation of cellular clusters causes the cell preparations to become inhomogeneous and this may alter the dynamics of excitation waves \cite{Bub:2002}. In particular, Steinberg \textit{et al.} reported that the propagation of excitation waves in a cardiac cell culture is affected by the concentration of collagen used to coat the substratum \cite{Benjamin:2006}.
Since we have observed that the formation of clusters is also affected by the concentration of collagen on the substratum, it is possible that the alternation of wave dynamics observed in Ref.~\cite{Benjamin:2006} is directly related to cluster formation.
Furthermore, while it has been demonstrated previously that the dynamics of excitation waves vary in a complicated manner over the time scale of several days \cite{Soen:1999, Hwang:2004, Kawahara:2002, Kawahara:2004}, the results presented here demonstrate that that formation of clusters is one of the essential factor to cause these differences in observed wave dynamics.

\section{Concluding Remarks}

In the present paper, the mechanism of spontaneous cluster formation in a primary culture of ventricular cells from neonatal rats was investigated.
On the basis of several observations, it is suggested that periodic contraction of cardiac muscle cells results in the relocation of focal contacts, and that this causes cells to aggregate.
In other words, the evolution of a cellular network is directly affected by the contraction activity of individual cells.

The mechanism described here is still hypothetical, and further elucidation is necessary.
By visualizing cytoskeletons and focal contacts in living cells using fluorescent proteins, more detailed information on the molecular processes taking place during cluster formation would be obtained.
Construction of theoretical models would also facilitate our understanding of the molecular mechanisms involved.

In addition to the molecular mechanism underlying cluster formation, it would be interesting to investigate how the formation of cluster structure affects the contraction dynamics of cells.
Moreover, the effect of spontaneous cell contraction on cellular properties, such as shape and the orientation of cells should also be investigated.

In previous studies, the contraction dynamics of cardiac cell cultures have been theoretically studied within the framework of conventional reaction-diffusion systems or coupled oscillators \cite{Bub:2002, Bub:2003, Soen:1999, Kawahara:2002, Kawahara:2004}, with age-dependent changes in the cellular network considered as merely changes in external parameters.
The present study, however, suggests that the changes in these parameters may in fact be induced by the dynamics of the contraction itself. It might therefore be possible that a new class of phenomena arises through interactions between the dynamics of contraction and the structure of the cellular network.
This research may therefore lead to the development of new strategies for controlling the architecture of cellular networks by modulating the contraction activity of cells.

\section*{Acknowledgments}
The authors thank Prof.~Y.~Atomi and Dr.~E.~Ohto at the University of Tokyo for technical advice regarding the preparation of primary cell culture, Mr.~S.~Watanabe at Kyoto University for experimental assistance related to microscopy, and Prof.~O.~Niwa at Kyoto University for experimental support. This work was supported by a grant from the Ministry of Education, Science, Sports and Culture of Japan, No.~18840021, and Inamori Foundation.

\end{document}